\documentclass{iopart}
\usepackage{graphicx}
\usepackage{iopams}
\newcommand{\url}{\tt}

\begin{document}

  \paper{Slowly Rotating Homogeneous Stars and the Heun Equation}
  
  \author{David Petroff}
  \address{Theoretisch-Physikalisches Institut, University of Jena,
            Max-Wien-Platz 1, 07743 Jena, Germany}
  \ead{D.Petroff@tpi.uni-jena.de}
  \begin{abstract}
   The scheme developed by Hartle for describing slowly rotating bodies in 1967
was applied to the simple model of constant density by Chandrasekhar 
and Miller in 1974.
The pivotal equation one has to solve turns out to be one of Heun's equations.
After a brief discussion of this equation and the chances of finding a closed
form solution, a quickly converging series solution of it is presented. A
comparison with numerical solutions of the full Einstein equations allows
one to truncate the series at an order appropriate to the slow rotation approximation.
The truncated solution is then used to provide explicit expressions for
the metric.

  \end{abstract}

% \submitto{\CQG}
% \maketitle
 
 \section{Introduction}
  Until now, no one has succeeded in finding an analytic solution to
Einstein's equations that describes an isolated, rotating, three
dimensional perfect fluid. If such a solution is to be found, then most
likely for a homogeneous body in equilibrium. After all, three limiting
cases are known analytically for this equation of state: the non-rotating
limit (inner and outer Schwarzschild solution), the portion of the
Newtonian limit made up of the Maclaurin spheroids and the disc limit (the
relativistic disc of dust, see \cite{NM95}).

Given such an analytic solution, it could then be expanded about various
limits. An appropriate post-Newtonian expansion would yield the Maclaurin
spheroids in the zeroth order, then the analytically known first
post-Newtonian corrections to the next order (see \cite{Chand67,
Bardeen71}) and so on. Such a post-Newtonian expansion can be expressed
entirely in terms of elementary functions at least up to the fourth order
\cite{P03}. Similarly, an expansion with respect to the angular velocity
would result in the inner and outer Schwarzschild solution in the zeroth
order. In this paper we shall consider the first order contribution to the
slow rotation
expansion using the formalism introduced by Hartle \cite{Hartle67}.

Chandrasekhar and Miller \cite{CM74} already considered this problem in
1974 by solving the equations numerically. Since there now exist computer
programs capable of solving the complete Einstein equations in the case of
stationarity and axial symmetry to extremely high accuracy, a further
numerical treatment of the slowly rotating approximation would 
only be useful inasmuch as it provides a fairly simple alternative for arriving
at a good approximation to the full field equations.
An analytic treatment, however, can pinpoint where the stumbling
blocks preventing further progress lie, and can augment the disc solution
\cite{NM95} in suggesting a ``lower bound'' for the complexity of the
full-solution. A discussion of some of the analytic issues that arise
in this context can be found in \cite{Perjes00} and references therein.

We present the basic equations for a slowly rotating homogeneous perfect
fluid in \S\ref{basic} and show that the equation of primary importance is
one of Heun's equations. In \S\ref{HE} we discuss some of the properties
of Heun's equations and consider what transformations could result in a
simpler equation. Various series solutions are discussed and derived in
\S\ref{series}, and use is made of the numerical solution of the full
Einstein equations in order to determine an appropriate truncation order.
Using these truncated series, approximate expressions for all metric
functions are discussed in \S\ref{approximate} and provided in
\ref{appendix}.
 
 \section{Basic Equations for Slow Rotation}\label{basic}
  As in \cite{CM74}, we devote this section to listing the fundamental
equations derived by Hartle and then specialize them to the case of
constant energy density.

The metric describing a slowly rotating, stationary and axisymmetric fluid
is given by
\begin{eqnarray}\label{metric}
% \begin{split}
\fl   ds^2 = -e^{2\nu_{\rm S}}
          \left(1 + 2h \right)\,dt^2  
           + e^{2\lambda_{\rm S}}  \left[  1
               + \frac{e^{2\lambda_{\rm S}}}{r}    
             2m  \right]\,dr^2 \nonumber \\
           + r^2\left(1+ 2k \right) \Bigl[ d\theta^2  
           + \sin^2\theta \left(d\varphi-\omega\,dt \right)^2
              \Bigr] +  {\cal{O}}(\Omega^3)
% \end{split}
\end{eqnarray}
with
\begin{eqnarray*}
 h =& h_0(r) + h_2(r)\,P_2(\cos\theta) \\
 m =& m_0(r) + m_2(r)P_2(\cos\theta) \\
 k =& k_2(r)P_2(\cos\theta) \\
 \omega =& \omega(r).
\end{eqnarray*}
In the above equations, $P_2(\cos\theta)$ denotes Legendre's polynomial of
order 2, $\Omega$ is the angular velocity (see \cite{Hartle67} for an
account of what `slow rotation' means in terms of $\Omega$), $\omega$ is of
order $\Omega$ and $h$, $m$ and $k$ are of order $\Omega^2$. In the above
expansion in spherical harmonics, use was made of a coordinate freedom in
order to transform away the term $k_0(r)$.

The non-rotating metric, obtained when $\Omega \to 0$, is
\begin{eqnarray*}
   ds^2 = -e^{2\nu_{\rm S}} \,dt^2  
           + e^{2\lambda_{\rm S}} \,dr^2
           + r^2 \left( d\theta^2 +
             \sin^2\theta\,d\varphi^2 \right).
\end{eqnarray*}
For a non-rotating perfect fluid with pressure $p_{\rm S}$, constant energy
density $\varepsilon$, radius%
\footnote{The coordinate radius of the star $r_{\rm S}$ is not to be confused
with $R_{\rm S}$ in \cite{CM74}, which they use to denote the Schwarzschild radius.}
$r_{\rm S}$ and mass
\begin{equation}\label{M_S}
M_{\rm S}=\frac{4}{3} \pi \varepsilon r_{\rm S}^3,
\end{equation}
the metric functions are%
\footnote{We adopt units in which $c=G=1$.}
(see e.g.\ \cite{Stephani04})
\begin{eqnarray}
   r \ge r_{\rm S}:  \qquad
  &e^{\lambda_{\rm S}} = \frac{r}{r-2M_{\rm S}},  \quad
  &e^{\nu_{\rm S}}     = \frac{r-2M_{\rm S}}{r} \\
   r \le r_{\rm S}:  \qquad
  &e^{\lambda_{\rm S}} = \frac{1}{B(r)},  \quad
  &e^{\nu_{\rm S}}     = \frac{1}{2} [A - B(r)] 
\end{eqnarray}
with
\begin{equation}\label{AB}
  A := 3\sqrt{1-\frac{2M_{\rm S}}{r_{\rm S}}}, \qquad 
  B(r) := \sqrt{1-\frac{2M_{\rm S}r^2}{r_{\rm S}^3}} 
\end{equation}
and the normalized pressure is given by
\[ \frac{p_{\rm S}}{\varepsilon} = \frac{A/3 - B}{B-A}.\]
Integrating the equation for hydrostatic equilibrium, it is natural to
define
\begin{eqnarray}
  p^* &=\, \ln(\varepsilon + p) - 
    \int^\varepsilon \frac{d\varepsilon'}{\varepsilon'+p(\varepsilon')}\\
    &=\, \ln(\varepsilon + p) \qquad (\varepsilon={\rm constant}), \nonumber
\end{eqnarray}
which we expand as with the metric coefficients
\[ p^* = p^*_{\rm S} + \delta p^*_0(r) + 
         \delta p^*_2(r)\,P_2(\cos \theta).\]
Hydrostatic equilibrium is then realised when
\begin{eqnarray}\label{pressure}
% \begin{split}
 \eqalign{
  \delta p^*_0 + h_0 -\frac{1}{3}r^2 e^{-2\nu_{\rm S}}\tilde \omega^2 =& \,
    {\rm constant} =\, \gamma \\
  \delta p^*_2 + h_2 + \frac{1}{3}r^2 e^{-2\nu_{\rm S}}\tilde \omega^2 =& \,
    0,} 
% \end{split}
\end{eqnarray}
where we have introduced%
\footnote{Our notation $\tilde\omega = \Omega -\omega$ corresponds to $\bar{\omega}$
in \cite{Hartle67} and $\varpi$ in \cite{CM74}. In this paper, a bar above a symbol
will denote a dimensionless quantity.}
\[\tilde\omega := \Omega -\omega.\]
Letting
\begin{eqnarray*} 
 j:=\,& e^{-(\lambda_{\rm S} + \nu_{\rm S})} \\
 v_2 :=\,& h_2 + k_2,
\end{eqnarray*}
and restricting ourselves to the case of constant density, we can write the
field equations as
%
%\begin{widetext}
%\begin{subequations}\label{fullsystem}
\numparts
 \begin{eqnarray}\label{fullsystem}
\fl   \frac{d^2 \tilde\omega}{dr^2} &=\,
       -\frac{d\tilde\omega}{dr}\frac{d}{dr}\ln(r^4j)
       - \frac{4\tilde\omega}{r}\frac{d}{dr}\ln(j) \label{om:orig}\\
\fl    \frac{dm_0}{dr} &=\, 
        \frac{1}{12}r^4j^2\left(\frac{d\tilde\omega}{dr}\right)^2
       - \frac{1}{3}r^3 \tilde\omega^2 \frac{d}{dr}j^2 \label{m0:orig}\\
% %  \begin{split}
\fl    \frac{dh_0}{dr} &=\, \left[ m_0 e^{2\lambda_{\rm S}}
         \left(\frac{1}{r^2} + 8\pi p_{\rm S} \right)
         - \frac{1}{12}r^3j^2
                   \left(\frac{d\tilde\omega}{dr}\right)^2
        + 4\pi r \left(\varepsilon+p_{\rm S}\right)\delta p^*_0
         \right] e^{2\lambda_{\rm S}}\label{h0:orig}\\
% %  \end{split} \\
% %  \begin{split}
\fl     \frac{dh_2}{dr} &=\, \left(\frac{-2 e^{2\lambda_{\rm S}}}{r^2}\Big/\frac{d\nu_{\rm S}}{dr}\right) v_2
                                +\left[-2\frac{d\nu_{\rm S}}{dr}
       - \frac{1}{r}\Big/ \frac{d\nu_{\rm S}}{dr}
                             \left(\frac{1}{2j^2}\frac{d\left(j^2\right)}{dr} 
                               + \frac{d\lambda_{\rm S}}{dr}\right) \right]\,h_2
  \nonumber \\\fl &\,
       +  \left[ \frac{r^2\,j^2}{6}\left(\frac{d\tilde\omega}{dr}\right)^2 
                 -\frac{r\,\tilde{\omega}^2}{3}\frac{d\left(j^2\right)}{dr}\right]
   \left(r^2\frac{d\nu_{\rm S}}{dr} 
             - \frac{e^{2\lambda_{\rm S}}}{2}\Big/\frac{d \nu_{\rm S}}{dr} \right) \label{h2:orig}\\
% %  \end{split} \\
% %  \begin{split}
\fl     \frac{dv_2}{dr} &=\, -2\frac{d\nu_{\rm S}}{dr} h_2 + \left(\frac{1}{r} 
        + \frac{d\nu_{\rm S}}{dr}\right)
                    \left[\frac{1}{6}r^4j^2\left(\frac{d\tilde\omega}{dr}\right)^2
                          - \frac{1}{3}r^3\tilde\omega^2\frac{d\left(j^2\right)}{dr} \right]
            \label{v2:orig}\\ 
% %  \end{split} \\
% %  \begin{split} 
\fl    m_2 &=\, r e^{-2\lambda_{\rm S}}
           \left(-h_2 +
           \frac{1}{6}r^4 j^2\left(\frac{d\tilde\omega}{dr}\right)^2
             -\frac{1}{3} r^3 \tilde\omega^2 \frac{d\left(j^2 \right)}{dr} \right)\label{m2:orig}.
% %  \end{split}
 \end{eqnarray}
 \endnumparts
%\end{subequations}
%\end{widetext}
 
In the vaccum region, where $\varepsilon=p=0$ and $j=1$, the solutions
to the above equations are
%
%\begin{subequations}
 \numparts
 \begin{eqnarray}
  r \ge r_{\rm S}: & \nonumber \\
  \tilde\omega =&\, \Omega- \frac{2J}{r^3} \label{omega_out}\\
  m_0 =&\, \delta M-\frac{J^2}{r^3} \label{m0_out}\\
  h_0 =&\, \frac{\delta M}{r-2M_{\rm S}} + \frac{J^2}{r^3(r-2M_{\rm S})} \\
  h_2 =&\, \frac{J^2}{r^3}\left(\frac{1}{M_{\rm S}}+\frac{1}{r} \right) +
           KQ_2^2\left(\frac{r}{M_{\rm S}} -1 \right) \label{h2:out} \\
  v_2 =&\, -\frac{J^2}{r^4} - K\frac{2M_{\rm S}}{\sqrt{r(r-2M_{\rm S})}}
           Q_2^1\left(\frac{r}{M_{\rm S}} -1 \right) \label{v2:out} \\
  m_2 =&\, \frac{1}{9}rA^2\left(-h_2 + \frac{6 J^2}{r^4} \right)         ,
 \end{eqnarray}
 \endnumparts
%\end{subequations}
%
where $Q_n^m$ is the associated Legendre function of the second kind%
\footnote{The minus sign in front of the second term of \eref{v2:out}
has to do with the normalization of the Legendre function and the choice
of the branch cut.}
and $J$, $\delta M$ and $K$ are constants, $J$ being the angular momentum
and $\delta M$ the change in mass with respect to the non-rotating configuration.

Equations~\eref{fullsystem} are listed in a hierarchical order, each
equation being soluble once the preceding equations have been solved
(note that \eref{h2:orig} and \eref{v2:orig} form a coupled system).
Eq.~\eref{om:orig} plays a pivotal role, since the solution
of the remaining equations relies on the solution of this equation. Hence
we shall devote the next section entirely to this equation. Before doing so,
it will be convenient to put \eref{om:orig}, which is a Heun equation,
into the standard form (see e.g.\ \cite{Ronveaux}).

To bring it into this form, let us begin with a transformation of variables
in \eref{om:orig} from $r$ to the $B(r)$ of \eref{AB}. Making
use of the constant $A$ from \eref{AB}, we find
\begin{equation}
 (B-A)(B^2-1)\frac{d^2\tilde\omega}{dB^2} 
  + (4B^2-5AB+1)\frac{d\tilde\omega}{dB} - 4A\tilde\omega =0.
\end{equation}
With the further substitution (see \cite{Kamke} (2.329))%
\footnote{Both this substitution and the resulting \eref{eq:Heun}
are not unique, since there is freedom as to how the original four
singularities of the equation, $(1,-1,A,\infty)$, are mapped onto 
$(0,1,a,\infty)$. This freedom will be considered amongst the
transformation of \S\ref{HE}.\label{uniqueness}}
\[z:=\frac{1-B}{2}\]
we arrive at
\begin{equation}\label{eq:Heun}
 \frac{d^2\tilde\omega}{dz^2} 
  + \left(\frac{\gamma}{z}+ \frac{\delta}{z-1} +\frac{\epsilon}{z-a}\right)
  \frac{d\tilde\omega}{dz} + \frac{\alpha\beta z -q}{z(z-1)(z-a)}\tilde\omega
   =0
\end{equation}
with
\begin{eqnarray*}
% \begin{split}
   &a = \frac{1-A}{2}, \quad q = -2A,
        \quad \alpha = 3, \quad \beta = 0, \quad \gamma = \frac{5}{2}, \\
   &\delta = \frac{5}{2}, \quad
  \epsilon = \alpha + \beta  +1 -\gamma - \delta \quad \Longrightarrow
   \epsilon = -1.
% \end{split}
\end{eqnarray*}

 \section{Heun's Equation: Properties and Transformations}\label{HE}
  Using the terminology of \cite{Ronveaux}, we consider three categories of
solutions to Heun's equation: {\sl local solutions}, {\sl Heun functions} and
{\sl Heun polynomials}.

{\bf Local solutions} to Heun's equation are valid in the
neighbourhood of one singularity and are associated with one of the two
exponents there. Since there are four singularities, there are a total of eight
such solutions. {\bf Heun functions} are solutions valid in an region containing two
adjacent singularities. {\bf Heun polynomials} are solutions to Heun's equation
valid at three singularities. Despite the name, not all such solutions are
polynomials, but are of the simple form
\begin{equation}\label{eq:HP}
  H\!p\,(z) = z^{\sigma_1}(z-1)^{\sigma_2}(z-a)^{\sigma_3}p_n(z),
\end{equation}
where $p_n(z)$ is a polynomial of degree $n$ and $\sigma_{1,2,3}$ is one of the
exponents associated with the singularity at $z=0,1,a$ respectively.

In order to determine which type of solution we need to look for, we have to
determine the domain of relevance for our particular Heun equation given by
\eref{eq:Heun}. The parameter $A$ and the quantities that depend on it are
defined on the intervals
\[ A\in (1,3), \qquad a \in (0,-1), \qquad q\in (0,-3/2). \]
The limit of inifinite central pressure is given by $A\to 1$ and the Newtonian
limit by $A \to 3$. We exclude these limits in this analysis (as indicated by
the open intervals), but note that Chandrasekhar and Miller include an interesting
discussion of the limit of infinite central pressure in \cite{CM74} and believe
that consideration of such limits in the context of the confluent Heun equation
could lead to interesting results.

The interior of the star, i.e.\ the region of validity of \eref{eq:Heun},
is given by the interval
\begin{equation}\label{z-interval}
 B\in [1,A/3] \Longrightarrow z\in \left[0,\frac{3-A}{6}\right]=
                                            \left[0,\frac{a+1}{3}\right],
\end{equation}
where $z=0$ represents the centre of the star and $z=(3-A)/6$ its surface.

Since $a$ is always strictly negative whereas $z$ is always real and positive,
we are interested in a solution including the singularity at zero and extending along
the real axis toward the singularity at $z=1$ (though not reaching it). The function
$\tilde\omega$ must remain finite (and non-zero) at the point $z=0$, whence
we require at the very least the local solution there corresponding to the exponent
$\sigma_1=0$.

The remainder of this section is devoted to a survey of attempts that
can be made to find a closed form solution to our Heun equation
satisfying these constraints. Although none of these attempts proved
successful, the following subsections are intended to stimulate new
ideas in this direction, explain where problems lie in what may seem
promising solution strategies and help the reader avoid choosing
fruitless paths of thought.

\subsection{Heun Polynomials}

The most tractable solution to Heun's equation is one
of the Heun polynomials, so that we begin by searching for such a solution even
though the validity at the two `additional' singularities is not necessary
for our purposes. As was mentioned in footnote~\ref{uniqueness}, the particular
form of the Heun equation given in \eref{eq:Heun} is not unique since various
transformations map a Heun equation onto a new Heun equation with other parameters
and exponents. Indeed there are 192 mappings of Heun's equation onto itself, corresponding to the 192 local solutions that can be provided for it. These are made up of 24 transformations of the independent variable, which map the four singularities onto themselves, and 8 elementary power transformations of the dependent variable. A chapter on these transformations is contained in \cite{Ronveaux} and a discussion of their group structure as well as a useful table containing all of them can be found in \cite{Maier04}. These transformations will not occupy us further in this paper, but the statements made and conclusions drawn are valid (modulo obvious modifications resulting from the transformation) for all mappings of \eref{eq:Heun} onto itself.

The exponents $\sigma_i$ (see \eref{eq:HP}) of the four Heun polynomials
consistent with our physical problem as well as the parameters $\alpha$ and $\beta$
for the solutions  are listed in Table~\ref{tab:HP}.
\begin{table}
  \caption{The values of the exponents $\sigma_i$ of \eref{eq:HP} and the
     parameters $\alpha$ and $\beta$ corresponding to the four classes of Heun
	  polymials that would yield permissible solutions to our physical problem. The
	  numbering of the classes follows the conventions in \cite{Ronveaux}.
	  \label{tab:HP}}
  \begin{indented}
  \item[]\begin{tabular}{cccccc}\br
   Class & $\sigma_1$ & $\sigma_2$ & $\sigma_3$ & $\alpha$ & $\beta$ \\ \mr 
  I     & 0  & 0          & 0     & $-n$           & $\gamma+\delta+\epsilon+n-1$ \\
  III   & 0  & $1-\delta$ & 0     & $\delta-n-1$   & $\gamma+\epsilon+n$ \\
  V     & 0  & 0   & $1-\epsilon$ & $\epsilon-n-1$ & $\gamma+\delta+n$ \\
  VII   & 0  & $1-\delta$ & $1-\epsilon$ & $\delta+\epsilon-n-2$ & $\gamma+n+1$ \\ \br  
  \end{tabular}
 \end{indented}
\end{table}
A comparison with \eref{eq:Heun} shows us that only a Heun polynomial of
class~I is consistent with the given parameters since $n$, the degree of the
polynomial $p_n(z)$ in \eref{eq:HP}, can of course only be a non-negative
integer. The set of equations that must be solved in general in order to determine
$p_n(z)$ provides a polynomial equation of degree $n+1$ for the accessory parameter
$q$. In our case, the equation is simply $q=0$, which lies just outside the allowed
range for $q$ and thus there exists no Heun polynomial, which is a solution of the
physical problem being considered.

\subsection{Heun Functions}

Since we require a solution valid at $z=0$ and extending along the real axis
toward $z=1$, a Heun Function seems to be the appropriate solution. It is
related to the two-point connection problem treated in \cite{SS80} and 
discussed further in \cite{SL00}. The search for such solutions generally
implies solving an eigenvalue problem and is valid only for a restricted set
of values for the accessory parameter $q$. In our case, $q$ is not independent
of the singularity $a$, and no such solutions exist.

\subsection{Rational and other Transformations}

The next step we take in looking for a solution of our Heun equation is to
consider a broader set of transformations. Since Heun's equations are closely
related to the hypergeometric equations, one can hope to find a transformation
relating the two equations. Kuiken \cite{Kuiken79} addressed the question as to
when a hypergeometric equation can be transformed into a Heun equation via a
rational substitution of the independent variable (excluding the trivial case
$\alpha \beta = q =0$). Her work was reexamined and completed by Maier
\cite{Maier02}. Using their results, we have found that no rational transformation of the independent
variable maps our Heun equation (with the specified parameter range) onto a
hypergeometric equation%
\footnote{Many of the computations in this paper made use of the computer algebra
 program Maple\texttrademark. Maple is a trademark of Waterloo Maple Inc.}.
Even had such a transformation been found, it would only have been valid for
specific values (or one specific value) of the parameter $A$, because of the
nature of the necessary conditions for such a transformation to exist.

Having ruled out rational transformations, we turn our attention briefly to
integral transformations derived by Carlitz and Valent and described in
\cite{Valent86}, which impose no restriction on the accessory parameter. Such
transformations have proved useful in finding previously unknown, closed form
solutions to Heun's equation. In our case, these methods only allow for marginal
progress. Consider first a transformation of the independent variable
\[ B' = \frac{B-a}{B(1-a)}.\]
Applying to this the transformation (9) of \cite{Valent86}, and denoting the new
quantities by a douple prime, one finds that
\[ \gamma''=\varepsilon''=\frac{1}{2} \qquad {\rm and} \qquad 
   \alpha''+\beta''=\delta''\]
and can apply the quadratic transformation discussed on pg.~59 of \cite{Erdelyi3}.
Here, however, one is not guaranteed that the solution will be regular at the centre
of the star.

More modern ideas, such as relating the Heun equation to the Schr\"odinger equation
through the ``generalized associated Lam\'e" (GAL) potential  seem
tailor made to the situation being considered here. Indeed, the analogue of (32)
in \cite{KS05} with $b=3/2$ can be applied to \eref{eq:Heun}, but the restriction
to the accessory parameter is $q=0$ as it was in the case of a Heun polynomial.

We know in fact that no meromorphic solution to our Heun equation exists (for arbitrary $q$)
since the necessary condition $\epsilon=1/2-m$, $m \in \mathbb{Z}$ \cite{Valent05} is not met.

 \section{Series Solutions}\label{series}
  The power series solution (Frobenius solution) is the most obvious
local solution to Heun's equation. It could be used to generate a
solution regular at the point $z=0$ and corresponding to the exponent zero,
\[ S := \sum_{m=0}^\infty \tilde{c}_m z^m,\]
which converges, in general, within
a circle extending out to the next singularity. It is also possible
to generate a series solution in terms of hypergeometric functions
as described in \cite{Erdelyi42}
%
%\begin{widetext}
\begin{eqnarray}\label{H_series} 
\fl H :=& \sum_{m=0}^\infty c_m \phi_m, \qquad {\rm with}\\
\fl \phi_m =& \frac{\Gamma(\alpha-\delta+m+1)\Gamma(\beta-\delta+m+1)}
                {\Gamma(\alpha+\beta-\delta+2m+1)}
          z^m F(\alpha+m,\beta+m; \alpha+\beta-\delta+2m+1;z),
         \nonumber
\end{eqnarray}
where $\Gamma(x)$ is the gamma function and $F(a,b\,;c\,;z)$ the
hypergeometric function. The coefficients $c_m$ are given by the
three term recurrence relation
\begin{eqnarray}\label{recursion}
% \begin{split}
   \eqalign{
    L_0 c_0  + M_0 c_1 =\, 0 \\
    K_m c_{m-1} + L_m c_m + M_m c_{m+1} =\, 0 \qquad (m=1, 2, 3, \ldots)}
% \end{split}
\end{eqnarray}
where
\begin{eqnarray*}
\fl K_{m+1} &=\, a \frac{(\alpha+m)(\beta+m)(\epsilon+m)(\alpha+\beta-\delta+m)}
                   {(\alpha+\beta-\delta+2m)(\alpha+\beta-\delta+2m+1)} \\
\fl L_m     &=\,  am(\gamma+m-1)\Bigl[\frac{(\alpha+m)(\alpha-\delta+m+1) + 
                                           (\beta+m)(\beta-\delta+m+1)}
                       {(\alpha+\beta-\delta+2m-1)(\alpha+\beta-\delta+2m+1)}\\
   \fl  &\,\qquad           - \frac{1}{\alpha+\beta-\delta+2m-1}\Bigl] 
            -m(\alpha+\beta-\delta+m) - q \\
   \fl  &\,\qquad            + a \frac{\alpha\beta(\gamma-2m) - \epsilon m(\delta-m-1)}
                        {\alpha+\beta-\delta+2m+1} \\
\fl M_{m-1} &=\, a \frac{(\alpha-\delta+m)(\beta-\delta+m)m(\gamma+m-1)}
                   {(\alpha+\beta-\delta+2m-1)(\alpha+\beta-\delta+2m)}.
\end{eqnarray*}
%\end{widetext}

The analysis in \cite{Erdelyi42} shows that $H$ converges in the domain
\begin{equation}
 \left|\frac{1-(1-z)^{1/2}}{1-(1+z)^{1/2}}\right| <
 \left|\frac{1-(1-a)^{1/2}}{1-(1+a)^{1/2}}\right|.
\end{equation}
Taking into account the range of $z$ (Eq.~\ref{z-interval}),
in turns out that both of these series will converge for
$A \in (3/2,3)$, but not in the range $A \in (1,3/2]$. If one chooses
to use this approximation as a rough model for neutron stars, one
is unlikely to be interested in $A<1.5$ however.%
\footnote{For a density of $\varepsilon=10^{18}\,{\rm kg}/{\rm m}^3$,
 $A=3/2$ corresponds in the static model to $M_{\rm S}\approx
 5.5\times 10^{30}\,{\rm kg}=2.8\, {\rm M}_\odot$.}

There are two reasons for opting for the series of hypergeometric
functions $H$ in favour of the power series $S$. For one thing, it generally
converges more quickly than the corresponding power series. In addition,
it is an alternating series in which successive partial sums straddle the
limiting value of the sum, so that they provide both an upper and lower
bound. A comparison of the two series can be found in
Table~\ref{tab:convergence} for the value of the function $\tilde{\omega}$
at the star's surface $z=(3-A)/6$ for $A=5/2$. The partial sums $H_n$ tend
to be about one order of magnitude more accurate than the corresponding
$S_n$. Taking into account that $H$ is an alternating series, a weighted
average such as the one presented in the last column of the table, ultilizes
$H_n$ and $H_{n+1}$ to produce a somewhat more accurate estimate for $H$ than
would be given by $H_{n+1}$ alone. The weight chosen in the table is based on
the fact that each term in the series (\ref{H_series}) is multiplied by $z^m$.

\begin{table}
\caption{The error of the partial sums $S_n$ and $H_n$ is shown at the point
         on the star's surface $z=(3-A)/6$ for the value $A=5/2$.\label{tab:convergence}}
 \begin{indented}
  \item[]\begin{tabular}{cccc}\br
    $n$ & $|S-S_n|$          & $|H-H_n|$          & $|H-\frac{1}{1+z}(z H_n+H_{n+1})|$ \\ \mr
    0   & $2 \times 10^{-1}$ & $2 \times 10^{-1}$ & $2 \times 10^{-2}$ \\
    1   & $2 \times 10^{-2}$ & $4 \times 10^{-3}$ & $2 \times 10^{-4}$ \\
    2   & $2 \times 10^{-3}$ & $1 \times 10^{-4}$ & $3 \times 10^{-6}$ \\
    3   & $2 \times 10^{-4}$ & $7 \times 10^{-6}$ & $6 \times 10^{-8}$ \\
    4   & $2 \times 10^{-5}$ & $5 \times 10^{-7}$ & $1 \times 10^{-9}$ \\
    5   & $2 \times 10^{-6}$ & $5 \times 10^{-8}$ & $5 \times 10^{-10}$\\ \br
  \end{tabular}
 \end{indented}
\end{table}

\begin{figure}
 \centerline{\includegraphics{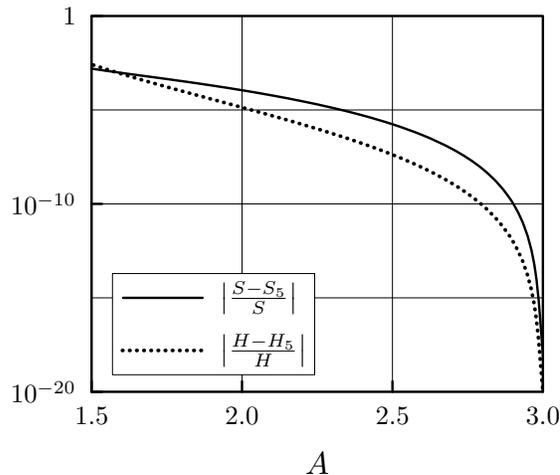}}
 \caption{$|(S-S_5)/S|$ and $|(H-H_5)/H|$ on the surface of the star $z=(3-A)/6$ are
  depicted over the range $A \in (3/2, 3)$.\label{fig:convergence}}
\end{figure}

Whereas the table shows that $H$ converges faster than $S$ for
$A=5/2$, Fig.~\ref{fig:convergence} shows how close $S_5$ and $H_5$ have
come to their asymptotic values for the full range of
convergent $A$ values. We see that both series converge very quickly for $A\to 3$
and that the hypergeometric series converges faster than the power
series for most of the convergent interval.

 \section{The Approximate Metric}\label{approximate}
  \subsection{Determining the Truncation Order}

Having found a convergent series solution to Heun's equation
(for a certain range of $A$), we can now compare the slow
rotation approximation with the solution of the full 
Einstein equations for axially symmetric, stationary, homogeneous,
uniformly rotating fluids to motivate how to choose the truncation
order of the partial sum $H_n$. The solution of Einstein's equations
can be found numerically, and using the spectral program described
in \cite{AKM3}, we can reach machine accuracy and thus have an
absolute measure for the accuracy of the Hartle approximation.
We choose to compare configurations with the same total mass%
\footnote{A discussion as to how one can integrate \eref{om:orig}
to determine the mass follows in \S~\ref{remaining}.}
$\bar{M}$ and angular velocity $\bar{\Omega}$,
where the bar indicates that the quantity is expressed in terms
of units of the energy density $\varepsilon$.

We are interested in solutions of Einstein's equations that are
connected to the non-rotating limit via a continuous parameter
transition. Such solutions were studied in \cite{SA} and termed
the `generalized Schwarzschild' class of solutions. A sequence
of stars from this class with constant mass
$\bar{M}< 4/9 \bar{R}_{\rm S} \approx 0.145$ can be followed from 
the non-rotating limit to the mass-shedding limit, at which a cusp
forms along the equatorial rim. Along any such sequence, $\bar{\Omega}$
reaches a global maximum somewhere between the two limiting points.
In the approximation considered here, however, the mass is monotonic
in $A\in(3/2,3)$ for a given $\bar{\Omega}$. Therefore it is not
possible to find more than one configuration for given $\bar{M}$ and
$\bar{\Omega}$ and we are forced to restrict our attention to the
portion of the sequence between the static limit and the maximum value
for $\bar{\Omega}$.

\begin{figure}
 \centerline{\includegraphics{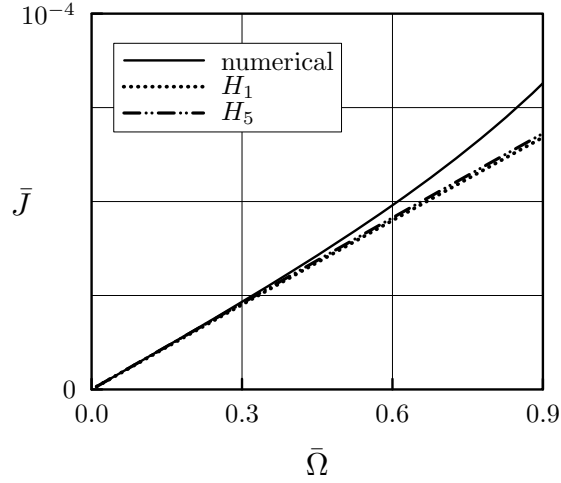}}
 \caption{The angular momentum $\bar J$ is shown as a function of angular
  velocity $\bar \Omega$ for a sequence of star's with the constant mass
  $\bar M = 0.01$. The solid line depicts numerical results, which are
  accurate to better than 10 digits and thus act as an absolute standard
  of reference.\label{M=0.01}}
\end{figure}

\begin{figure}
 \centerline{\includegraphics{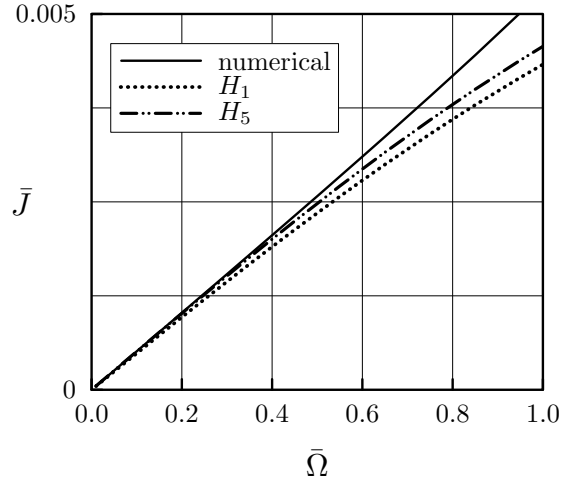}}
 \caption{As in Fig.~\ref{M=0.01}, but for a sequence with the mass
 $\bar M=0.1$.\label{M=0.1}}
\end{figure}

In Figs~\ref{M=0.01} and \ref{M=0.1} we see how $\bar J$ depends on
$\bar\Omega$ for one sequence close to the Newtonian limit ($\bar M=0.01$)
and one highly relativistic configuration ($\bar M=0.1$). We see verified
in each plot that the approximation becomes arbitrarily good for
$\Omega \to 0$. Furthermore, we see in agreement with
Fig.~\ref{fig:convergence} that $H$ converges quickly for large $A$, and
indeed $H_5$ is bairly distiguishable from $H_1$ in this figure ($A\approx 2.8$
over the whole range of the plot). This does not mean that the slow rotation
approximation is good however. The discrepancy between the correct and approximated
values for $\bar J$ are quite noticable for $\bar \Omega=0.6$, although its maximal
value along this sequence is $\bar \Omega \approx 1.23$. The inaccuracies due to
the slow rotation approximation are evident in Fig.~\ref{M=0.1} as well. Here,
however, the improvement brought about in going from $H_1$ to $H_5$ is more
pronounced ($A$ varies from about $2.7$ to $2.8$ over the range of the plot).  

To have more quantitative information as to when to truncate the series $H$,
Tables~\ref{tab:M=0.01} and \ref{tab:M=0.1} consider the same sequences and
show the relative error in the angular momentum as a function of $H_n$ for
various values of $\bar \Omega$. Each table shows a range of $\bar \Omega$
values extending up to its maximum for the respective sequence. The extremely
large errors emphasize how poor the slow rotation approximation is for
homogeneous bodies near the mass-shedding limit. One should keep in mind however
that these errors are not astrophysically relevant since the fastest known
pulsars have an angular velocity of $\bar\Omega\approx 0.4$, meaning that
the approximation holds to within about 3\%. If one
chooses the order of truncation, by requiring that the relative error in
$\bar J$ be in the vicinity of a few percent whenever the slow rotation
approximation allows for such accuracy, then $H_2$ seems an appropriate
choice. When providing explicit expressions for metric functions in the
next section, we shall choose this truncation order.

\begin{table}
\caption{The relative error in the angular momentum as a function of $H_n$ is shown for
   various values of $\bar \Omega$ for a sequence with $M=0.01$\label{tab:M=0.01}}
 \begin{indented}
  \item[]\begin{tabular}{lcccc}\br
    $\bar{\Omega}$ & $\left|\frac{J-J(H_1)}{J}\right|$  & $\left|\frac{J-J(H_2)}{J}\right|$ 
        & $\left|\frac{J-J(H_3)}{J}\right|$ & $\left|\frac{J-J(H_\infty)}{J}\right|$  \\ \mr
    0.01 &  $1.5 \times 10^{-2}$ & $2.5 \times 10^{-4}$ & $2.6\times 10^{-5}$ & $1.8\times 10^{-5}$ \\
    0.1  &  $1.7 \times 10^{-2}$ & $1.6 \times 10^{-3}$ & $1.8\times 10^{-3}$ & $1.8\times 10^{-3}$ \\
    0.2  &  $2.3 \times 10^{-2}$ & $7.1 \times 10^{-3}$ & $7.4\times 10^{-3}$ & $7.3\times 10^{-3}$ \\
    0.4  &  $4.5 \times 10^{-2}$ & $2.9 \times 10^{-2}$ & $3.0\times 10^{-2}$ & $3.0\times 10^{-2}$ \\
    0.8  &  $1.4 \times 10^{-1}$ & $1.3 \times 10^{-1}$ & $1.3\times 10^{-1}$ & $1.3\times 10^{-1}$ \\
    1.23 &  $4.7 \times 10^{-1}$ & $4.6 \times 10^{-1}$ & $4.6\times 10^{-1}$ & $4.6\times 10^{-1}$ \\ \br
  \end{tabular}
 \end{indented}
\end{table}

\begin{table}
\caption{As in Table~\ref{tab:M=0.01}, but for a sequence with $M=0.1$.\label{tab:M=0.1}}
 \begin{indented}
  \item[]\begin{tabular}{lcccc} \br
    $\bar{\Omega}$ & $\left|\frac{J-J(H_1)}{J}\right|$  & $\left|\frac{J-J(H_2)}{J}\right|$ 
        & $\left|\frac{J-J(H_3)}{J}\right|$ & $\left|\frac{J-J(H_\infty)}{J}\right|$  \\ \mr
    0.01 &  $5.2 \times 10^{-2}$ & $1.3 \times 10^{-2}$ & $5.8\times 10^{-3}$ & $1.5\times 10^{-5}$ \\
    0.1  &  $5.3 \times 10^{-2}$ & $1.1 \times 10^{-2}$ & $7.3\times 10^{-3}$ & $1.5\times 10^{-3}$ \\
    0.2  &  $5.8 \times 10^{-2}$ & $6.7 \times 10^{-3}$ & $1.2\times 10^{-2}$ & $6.0\times 10^{-3}$ \\
    0.4  &  $7.5 \times 10^{-2}$ & $1.1 \times 10^{-2}$ & $2.9\times 10^{-2}$ & $2.4\times 10^{-2}$ \\
    0.8  &  $1.4 \times 10^{-1}$ & $7.9 \times 10^{-2}$ & $9.5\times 10^{-2}$ & $9.0\times 10^{-2}$ \\
    1.45 &  $4.0 \times 10^{-1}$ & $3.6 \times 10^{-1}$ & $3.7\times 10^{-1}$ & $3.7\times 10^{-1}$ \\ \br
  \end{tabular}
 \end{indented}
\end{table}

\subsection{Deriving Explicit Expressions}\label{remaining}

If we label the arguments in the hypergeometric functions of $H$
(see \eref{H_series}) by $a,b$ and $c$, then we find
$a+b-c=-3/2$. By applying Gauss' relations for contiguous
hypergeometric functions (see e.g.\ \S~2.8 in \cite{Erdelyi1}),
we can convert this into two functions with $c=-1/2$ and $c=1/2$.
Quadratic transformations are known to exist for such functions
so that they can be converted into associated Legendre functions
of the form $P_\nu^{\pm n + 1/2}$, $n \in \mathbb Z$ (see e.g.\
\S 7.3.1 Eqs~(36) and 41 in \cite{Prudnikov3}). Associated Legendre
functions of this form can, in turn, be represented with a finite
number of terms, \S~3.6.1 in \cite{Erdelyi1}. The explicit expression
that results for $H_2$ can be found in the appendix.

The requirement that the metric functions be continuous at the
star's boundary leads via \eref{omega_out} to
\begin{equation}
 \left.\bar{\tilde\omega}\right|_{r=\bar{R}_{\rm S}} 
   = \bar{\Omega} - \frac{2\bar J}{\bar{R}_{\rm S}^3},
\end{equation}
where we have chosen to express the relation in terms
of dimensionless quantities. Taking into account
\begin{equation} \bar{M}_{\rm S}=\frac{4}{3} \pi \bar{R}_{\rm S}^3
\Longleftrightarrow \bar{R}_{\rm S}^2 = \frac{3}{8\pi}(1-A^2/9),
\end{equation}
(cf.\ \eref{M_S}), we thus know $\bar J$ as a function
of $c_0$ and $A$ for a given $\bar{\Omega}$. Since the normal
derivative of $\bar{\tilde\omega}$ must be continous on this 
surface (see e.g.\ \S~30.5 in \cite{Stephani04}),
\begin{equation}
 \left. \frac{d}{dr}\bar{\tilde\omega}\right|_{r=r_{\rm S}} 
   = \frac{6\bar J}{\bar{R}_{\rm S}^4}
\end{equation}
must hold as well. If we were to prescribe $A$ as in \cite{CM74}, then
this would suffice to determine $c_0$ and $\bar J$.

If on the other hand, we choose to prescribe $\bar{\Omega}$ and
$\bar M$ as in the previous section, then we have to solve
\eref{m0:orig} before being able to determine both $A$ and
$c_0$.

Eq.~\eref{m0:orig} can then be integrated from the centre out to the
surface of the star to give an
expression for $\delta\bar{M}= \bar{m}_0|_{r=\bar{R}_{\rm S}} +
\bar{J}^2/\bar{R}_{\rm S}^3$ from \eref{m0_out}. The constant
of integration must be chosen such that $m_0(0)=0$ so that $g_{rr}$
remains finite at the star's centre. The total
mass of the system is then $\bar{M} = \bar{M}_{\rm S} +\delta\bar{M}=
4/3 \pi \bar{R}_{\rm S}^3 +\delta\bar{M}$. Taking into account
our choice of normalization (cf.\ \eref{M_S})
\[ \bar{M}_{\rm S}=\frac{4}{3} \pi \bar{R}_{\rm S}^3
\Longleftrightarrow \bar{R}_{\rm S}^2 = \frac{3}{8\pi}(1-A^2/9),\]
we have arrived at expressions containing the unknown variables
$c_0$ and $A$, which can then be determined by prescribing $\bar{M}$ and
$\bar{\Omega}$.

Because the integrand in the integral to determine $m_0$ contains quadratic
terms in $\tilde\omega$ and its derivative, it does not seem possible to
obtain an analytic expression for $m_0$ based on the hypergeometric
representation discussed above. In the appendix, expressions are thus provided
for the metric functions in terms of a series in $z$.

Eq.~\eref{h0:orig} for determining $h_0$ is simply a first order, linear
differential equation, where \eref{pressure} is used to replace
$\delta p_0^*$ by an expression in $h_0$. The constant of  intergation can
be chosen arbitrarily%
\footnote{The comment on pg.~67 of \cite{CM74} that $h_0$ must vanish at
the origin seems to be an oversight. In that work, as here, $\delta p_0^*$
was chosen to vanish at the origin, which then implies through 
\eref{pressure} that $h_0(0)=\gamma$, which is not zero in general.}
and is chosen in the appendix such that $h_0(0)=\gamma$. This choice amounts
to identifying a rotating body with a non-rotating one of the same central
pressure since the choice implies through \eref{pressure} that
$\delta p_0^*(0)=0$. The nature of such choices of identification and some
of the effects different choices can have was discussed in the context of
the post-Newtonian approximation in \cite{P03}. The interior solution for
$h_0$ must join continuously to its exterior solution, thus fixing the
constant $\gamma$ of \eref{pressure}.

The system of equations \eref{h2:orig} and \eref{v2:orig} for obtaining
$h_2$ and $v_2$ can be solved as described in \cite{CM74}. Both functions
must vanish at the origin so that $h$ and $k$ have unique values there. The
remaining constant as well as the constant $K$ in \eref{h2:out} and
\eref{v2:out} can then be determined by requiring that these two functions
be continuous at the star's boundary. 
 
  \ack
   Thank you to Reinhard Meinel for suggesting this topic to me and for many helpful
   discussions. This work was supported by the Deutsche For\-schungs\-ge\-mein\-schaft
   (DFG) through the SFB/TR7 ``Gra\-vi\-ta\-ti\-ons\-wel\-len\-astro\-nomie'' and by a
   ``Lan\-des\-gra\-du\-ier\-ten\-sti\-pen\-di\-um'' from Thuringia. 
 
 \appendix
 
 \section{Explicit Representation of the Truncated Metric}\label{appendix}
  The formul\ae\ for the metric functions are collected in this appendix.
They are valid over the range $A \in (3/2,3)$ and make use of the variable
\[ z= \frac{1}{2}\left(1 - \sqrt{1-r^2/k^2}\right), \qquad 
         k=\frac{r_{\rm S}}{\sqrt{1-A^2/9}}.\]
The symbol $F(a,b;c;z)$ refers to the hypergeometric function. The metric
depends on three paramters (one scaling parameter and two ``physical''
parameters) that were here chosen to be $A=3\sqrt{1-2 M_{\rm S}/r_{\rm S}}$
(see \eref{AB}), $r_{\rm S}$ and $c_0$. $M_{\rm S}$ and $r_{\rm S}$ are the mass
and radius (in Schwarzschild coordinates) of a spherical star and $c_0$
is an integration constant (see \eref{H_series} and \eref{recursion})
giving the value of $\tilde\omega$ at the star's centre
$\tilde\omega(0)=4\sqrt{\pi}/3 c_0$. The expressions below do not diverge as
one approaches the Newtonian limit $A\to 3$ since $c_0 \to 0$ sufficiently
quickly in that limit.

\numparts
\begin{equation}
\fl \bar{\tilde \omega} =\, \frac{4\sqrt{\pi} \,c_0}{3} \left[1 +
              \frac{8A\,z}{5(A-1)}\,F\left(1,4;\frac{7}{2};z\right)
             + \frac{8(A\,z)^2}{35(A-1)^2}\,F\left(2,5;\frac{11}{2};z\right)\right] +\ldots
\end{equation}

\noindent which can be evaluated to yield
\begin{eqnarray}
\fl  \bar{\tilde \omega} =&\,\frac{\sqrt{\pi} c_0}{((A-1)\,z)^2 (z-1)}\Biggl[ \left( -1/24\,{\frac { \left( 88\,z-63 \right) {A}^{2}}{\sqrt {1-z}
\sqrt {z}}}+2/3\,{\frac {\sqrt {z}A}{\sqrt {1-z}}} \right) \arcsin\left( \sqrt {z} \right) \nonumber \\
\fl &\,  + \left( -{\frac {2}{45}}\,{z}^{3}-{\frac {
13}{45}}\,{z}^{2}+{\frac {23}{12}}\,z-{\frac {21}{8}} \right) {A}^{2}\nonumber \\
\fl &\, -2/9\,z \left( 4\,{z}^{2}-10\,z+3 \right) A+4/3\,{z}^{2} \left( z-1
 \right) 
 \Biggr] +\ldots
\end{eqnarray}
or expanding as a power series and dropping higher terms
\begin{eqnarray}
\fl \bar{\tilde \omega} =\,\frac{4\sqrt{\pi} \,c_0}{3} \left[1+ \frac{8A}{5(A-1)}z + 
                         \frac{8A(9A-8)}{35(A-1)^2}z^2 \right] + \ldots
\end{eqnarray}
\endnumparts
We now leave off the ellipsis at the end of each expression.
\begin{eqnarray}
\fl   m_0 =\, %{\frac {512}{525}}\,
\frac {512\,\pi\,{c_{{0}}}^{2}{r_{\rm S}}^{3} A\, {z}^{5/2}}
{ 525\left( 9-{A}^{2} \right) ^{3/2} \left( A-1 \right) ^{5}}
\,\biggl(
     840\, \left( A-1 \right) ^{2}
 +12\, \left( A-1 \right)  \left( A-125 \right) z \nonumber \\
 + \left( 2975+602\,A-441\,{A}^{2} \right) {z}^{2}
\biggr)
\\ \nonumber \\
\fl   \delta M =\, \frac {16\,\pi \,{c_{0}}^{2}{r_{\rm S}}^{3} \left( 3-A \right) ^{5/2}}
{127575 \left( 9-{A}^{2} \right) ^{3/2} \left( A-1 \right) ^{5} }
\times\nonumber \\
   \biggl(\sqrt {6}\,A\, \left(84015-109128\,A+34922\,{A}^{2}
+3176\,{A}^{3}-441\,{A}^{4}\right) \nonumber \\
 +{\frac {8}{63}}\, \left( 9-{A}^{2} \right)  \left( 
A-1 \right)  \left( A+3 \right) ^{5/2} \left( 45-56\,A+9\,{A}^{2}
 \right) ^{2}
\biggr)
\\ \nonumber \\
 \fl   h_0 =\, \gamma
-\frac {1792 A\,\pi \,{c_0}^{2}{{\it r_{\rm S}}}^{2}}
{15 \left( A-1 \right) ^{3} \left( {A}^{2}-9\right) }{z}^{2}

\\ \nonumber \\
\fl {\rm with} \nonumber \\
\fl   \gamma =\, \frac {-9\, \delta M}{{A}^{2} r_{\rm S}}
 + %{\frac {64}{893025}}\,
\frac {64 \pi\, {c_{0}}^2 {r_{\rm S}}^{2} \left( A-3 \right) }
{893025 {A}^{2} \left( A+3 \right)  \left( A-1\right) ^{4}}
 \nonumber \\
            \, \times \bigl(81\,{A}^{8}-522\,{A}^{7}-2102\,{A}^{6}+14262\,{A}^{5}+65961\,{A}^{4}\nonumber \\
-165591\,{A}^{3}-47466\,{A}^{2}+298890\,A-164025
\bigr) \nonumber \\ \nonumber \\
\fl   h_2 = \, f_0\,z 
+ \left( {\frac { \left( 3\,A-25 \right) f_0}{7(A-1)}}
+
%{\frac {512}{3}}\,
{\frac {512 \pi {c_0}^2 \,{r_{\rm S}}^{2}A\,}{3 \left( A^2-9 \right)  \left( A-1
 \right) ^{3}}} \right) {z}^{2}\nonumber \\ \hspace{-1cm}
+\left( {\frac {2 \left( 5\,{A}^{2}-17\,A+104 \right) f_0}{21 \left( A-1 \right) ^{2}}}
+
%{\frac {512}{225}}
\,{\frac {512\pi {c_0}^2 \,{r_{\rm S}}^{2} A\, \left( 129\,A-575 \right) }
{225 \left( A^2-9 \right)  \left( A-1 \right) ^{4}}}
 \right) {z}^{3}

\\ \nonumber \\
\fl   v_2 = \, \left( {\frac {-2 f_0}{A-1}}- 
{\frac {512\pi\,{c_0}^2{r_{\rm S}}^{2}A}{3 \left( A^2-9 \right)  
\left( A-1\right) ^{3}}} \right) {z}^{2}\nonumber \\
- \left({\frac {4 \left( A-13\right) f_0}{7 \left( A-1 \right) ^{2}}}
+ {\frac {1024\pi {c_0}^{2}\,{r_{\rm S}}^{2}A\, \left( 7A+25
 \right) }{75   \left( A^2-9 \right)  \left( A-1\right) ^{4}}} \right) {z}^{3}
\nonumber \\
+ \Biggl( {\frac {-2 \left( 5\,{A}^{2}-26\,A+221 \right) f_0}
{ 21\left( A-1 \right) ^{3}}} \nonumber \\
+ {\frac {512\pi {c_0}^2 \,{r_{\rm S}}^{2}A\, \left( 51\,{A}^{2}-2262\,A+9275
 \right) }{1575  \left( A^2-9 \right)  \left(A -1
 \right) ^{5}}} \Biggr) {z}^{4}

\end{eqnarray}

\noindent where the continuity of $h_2$ and $v_2$ at the star's boundary then give for the constant
$f_0$ and $K$ of \eref{h2:out} and \eref{v2:out}

\begin{eqnarray}
\fl  f_0 &= \, \pi {c_0}^{2} {Rs_{\rm S}}^{2}
\times
         \Biggl[-512\, \left( A-1 \right)  \left( A^2-9 \right)
 \Bigl( 81\,{A}^{8}-522\,{A}^{7}-3560\,{A}^{6}+23658\,{A}^{5}\nonumber \\
\fl & +44370\,{A}^{4}-120771\,{A}^{3}
-2046924\,{A}^{2}+3774195\,A-492075 \Bigr) Q_2^1\left(\frac{r_{\rm S}}{M_{\rm S}} -1 \right)\nonumber \\
\fl & -768\,A\, \Bigl(162\,{A}^{9}- 1206\,{A}^{8}-3160\,{A}^{7}+32728\,{A}^{6}
+14001\,{A}^{5}-328536\,{A}^{4}\nonumber \\
\fl &  +373590\,{A}^{3} -2968596\,{A}^{2}+4333095\,A + 328050\Bigr) Q_2^2\left(\frac{r_{\rm S}}{M_{\rm S}} -1 \right)
\Biggr] \Bigg/ \nonumber \\
         \fl & \Biggl[28350\, \left(A -3 \right)  \left( A+3 \right) ^{2} \left(A -1 \right) ^{3}
\nonumber \\
\fl & \qquad \times \left( 1989-2514\,A+962\,{A}^{2}-74\,{A}^{3}+5\,{A}^{4}
 \right) Q_2^1\left(\frac{r_{\rm S}}{M_{\rm S}} -1 \right)\nonumber \\
\fl & -14175\, \left( A-3 \right)  \left( A+3 \right) 
 \left( A-1 \right) ^{2}A\nonumber \\
\fl & \qquad \left(5\,{A}^{4} -92\,{A}^{3} +1790\,{A}^{2} -5052\,A+4149 \right)
 Q_2^2\left(\frac{r_{\rm S}}{M_{\rm S}} -1 \right)
\Biggr]
\\ \fl \nonumber \\
\fl K &=  {\frac {\pi {c_0}^{2} {r_{\rm S}}^{3}\, A }{8037225 r_{\rm S}\, \left( A-1 \right) ^{4}
 \left( A+3 \right) }}
\times
         \Biggl[64\, \Bigl( 2835\,{A}^{13}-65331\,{A}^{12}+893264\,{A}^{11}\nonumber \\
\fl & -4933304\,{A}^{10}-8773582\,{A}^{9}
+172596150\,{A}^{8}-361180413\,{A}^{7}-
958944555\,{A}^{6}\nonumber \\
\fl & +1549090818\,{A}^{5}+11997038790\,{A}^{4} -41313070791\,{A}^{3} +54020240631\,{A}^{2}\nonumber \\
\fl & -33095488275\,A+8082331875 \Bigr) 
\Biggr] \Bigg/ \nonumber \\
      \fl   & \Biggl[2\, \left( A-1 \right)  \left( A+3 \right)  \left( 5\,{A}^{4}-74\,{A}
^{3}+962\,{A}^{2}-2514\,A+1989 \right) Q_2^1\left(\frac{r_{\rm S}}{M_{\rm S}} -1 \right)\nonumber \\
\fl & -A\, \left( 5\,{A}^{4}-
92\,{A}^{3}+1790\,{A}^{2}-5052\,A+4149 \right) Q_2^2\left(\frac{r_{\rm S}}{M_{\rm S}} -1 \right)
\Biggr]
\end{eqnarray}
$m_2$ can then be calculated algebraically using \eref{m2:orig}.

\section*{References}
 
 \bibliographystyle{unsrt}
 \bibliography{Reflink}

\begin{thebibliography}{10}

\bibitem{NM95}
G.~{Neugebauer} and R.~{Meinel}.
\newblock General relativistic gravitational field of a rigidly rotating disk
  of dust: Solution in terms of ultraelliptic functions.
\newblock {\em Phys.\ Rev.\ Lett.}, 75:3046--3047, October 1995.

\bibitem{Chand67}
S.~Chandrasekhar.
\newblock The post-newtonian effects of general relativity on the equilibrium
  of uniformly rotating bodies ii. the deformed figures of the maclaurin
  spheroids.
\newblock {\em Astrophys.\ J.}, 147:334, 1967.

\bibitem{Bardeen71}
J.~M. Bardeen.
\newblock A reexamination of the post-newtonian maclaurin spheroids.
\newblock {\em Astrophys.\ J.}, 167:425, 1971.

\bibitem{P03}
D.~Petroff.
\newblock Post-newtonian maclaurin spheroids to arbitrary order.
\newblock {\em Phys.\ Rev.\ D}, 68:104029, 2003.

\bibitem{Hartle67}
J.~Hartle.
\newblock Slowly rotating relativistic stars i. equations of structure.
\newblock {\em Astrophys.\ J.}, 150:1005, 1967.

\bibitem{CM74}
S.~Chandrasekhar and J.~C. Miller.
\newblock On slowly rotating homogeneous masses in general relativity.
\newblock {\em Mon.\ Not.\ R.\ Astron.\ Soc.}, 167:63, 1974.

\bibitem{Perjes00}
Z.~Perj{\'e}s.
\newblock Rotating perfect fluid models in general relativity.
\newblock {\em Ann.\ Phys.\ (Leipzig)}, 9:368, 2000.

\bibitem{Stephani04}
H.~Stephani.
\newblock {\em Relativity}.
\newblock Cambridge University Press, Cambridge, 2004.

\bibitem{Ronveaux}
A.~Ronveaux, editor.
\newblock {\em Heun's Differential Equations}.
\newblock University Press, Oxford, Oxford, 1995.

\bibitem{Kamke}
E.~Kamke.
\newblock {\em Differentialgleichungen L\"osungsmethoden und L\"osungen}.
\newblock Akademische Verlagsgesellschaft, Leipzig, 1956.

\bibitem{Maier04}
R.~S. Maier.
\newblock The 192 solutions of the heun equation.
\newblock {\em math.CA/0408317}, 2004.

\bibitem{SS80}
R.~Sch\"afke and D.~Schmidt.
\newblock The connection problem for general linear ordinary differential
  equations at two regular singular points with applications in the theory of
  special functions.
\newblock {\em SIAM J.\ Math.\ Anal.}, 11(5):848--862, 1980.

\bibitem{SL00}
S.~Yu. Slavyanov and W.~Lay, editors.
\newblock {\em Special Functions: A Unified Theory Based on Singularities}.
\newblock University Press, Oxford, Oxford, 2000.

\bibitem{Kuiken79}
K.~Kuiken.
\newblock Heun's equation and the hypergeometric equation.
\newblock {\em SIAM J.\ Math.\ Anal.}, 10:655, 1979.

\bibitem{Maier02}
R.~S. Maier.
\newblock On reducing the heun equation to the hypergeometric equation.
\newblock {\em J.\ Differential Equations}, 213:171, 2005.

\bibitem{Valent86}
G.~Valent.
\newblock An integral transformation involving {H}eun functions and a related
  eigenvalue problem.
\newblock {\em SIAM J.\ Math.\ Anal.}, 17:688, 1986.

\bibitem{Erdelyi3}
A.~Erd{\'e}lyi, W.~Magnus, F.~Oberhettinger, and F.~G. Tricomi.
\newblock {\em Higher Transcendental Functions}, volume~3.
\newblock McGraw-Hill Book Company, New York, 1955.

\bibitem{KS05}
A.~Khare and U.~Sukhatme.
\newblock Quasi-periodic solutions of heun's equation.
\newblock {\em math-ph/0505077}, 2005.

\bibitem{Valent05}
G.~Valent.
\newblock {H}eun functions versus elliptic functions.
\newblock {\em math-ph/0512006}, 2005.

\bibitem{Erdelyi42}
A.~Erd{\'e}lyi.
\newblock The {F}uchsian equation of second order with four singularities.
\newblock {\em Duke Math.\ J.}, 9:48, 1942.

\bibitem{AKM3}
M.~Ansorg, A.~Kleinw{\"a}chter, and R.~Meinel.
\newblock Highly accurate calculation of rotating neutron stars: Detailed
  description of the numerical methods.
\newblock {\em Astron.\ Astrophys.}, 405:711, 2003.

\bibitem{SA}
K.~Sch{\"o}bel and M.~Ansorg.
\newblock Maximal mass of uniformly rotating homogeneous stars in einsteinian
  gravity.
\newblock {\em Astron.\ Astrophys.}, 405:405, 2003.

\bibitem{Erdelyi1}
A.~Erd{\'e}lyi, W.~Magnus, F.~Oberhettinger, and F.~G. Tricomi.
\newblock {\em Higher Transcendental Functions}, volume~1.
\newblock McGraw-Hill Book Company, New York, 1955.

\bibitem{Prudnikov3}
A.~P. Prudnikov, Yu.~A. Brychkov, and O.~I. Marichev.
\newblock {\em Integrals and Series}, volume~1.
\newblock Gordon and Breach Science Publishers, New York, 1990.

\end{thebibliography}
  
\end{document}